\documentclass[aps,prd,a4paper,preprintnumbers,twocolumn,floatfix,showpacs,nofootinbib]{revtex4}

\usepackage{psfrag}
\usepackage{subfigure}
\usepackage{color}
\usepackage{mathrsfs}
\usepackage{graphicx}

\def\be{\begin{equation}}
\def\ee{\end{equation}}
\def\bea{\begin{eqnarray}}
\def\eea{\end{eqnarray}}
\def\l{\left}
\def\r{\right}

\begin{document}


\title{Dynamical apparent horizons in inhomogeneous Brans--Dicke universes}


\author{Valerio Faraoni}
\affiliation{~Physics Department and {\em STAR} Research Cluster, 
Bishop's University
Sherbrooke, Qu\'ebec, Canada J1M~1Z7
}

\author{Vincenzo Vitagliano}
\affiliation{~CENTRA, Departamento de F\'isica, Instituto Superior T\'ecnico,
Universidade T\'ecnica de Lisboa - UTL, Av. Rovisco Pais 1, 1049 Lisboa, Portugal.
}

\author{Thomas P. Sotiriou}
\affiliation{~SISSA-International School for Advanced Studies, Via Bonomea
265, 34136 Trieste, Italy and  INFN, Sezione di
Trieste
}

\author{Stefano Liberati}
\affiliation{~SISSA-International School for Advanced Studies, Via Bonomea
265, 34136 Trieste, Italy and  INFN, Sezione di
Trieste
}



\begin{abstract} 
The presence and evolution of apparent horizons in a 
two-parameter family of  spherically symmetric, 
time-dependent solutions of Brans--Dicke 
gravity are analyzed. These solutions were introduced to 
model space- and time-varying gravitational 
couplings and are supposed to represent 
central objects embedded in a spatially flat universe. We 
find that the solutions possess multiple evolving apparent 
horizons, both 
black hole horizons covering a
 central singularity and cosmological ones. It is not uncommon for two 
of these horizons to merge, leaving behind a naked singularity covered 
only by a cosmological horizon. Two characteristic limits are also 
explicitly worked out: the limit where the theory reduces to general 
relativity and the limit where the solutions become static. The physical 
relevance of this family of solutions is discussed. 
\end{abstract}


\maketitle



\section{Introduction}

Varying ``constants'' of nature, first hypothesized by Dirac 
\cite{Dirac}, can be implemented naturally in the context of 
scalar-tensor gravity, in which the gravitational coupling 
becomes  a function of the spacetime point \cite{BransDicke61, 
ST}. String theories \cite{GreenSchwarzWitten}  
contain a  dilaton field coupling non-minimally to gravity which 
mimics a Brans--Dicke-like scalar field (indeed, it is well known 
that the low-energy limit of the bosonic string theory is an 
$\omega_0=-1$ Brans--Dicke theory \cite{bosonic}). Scalar-tensor 
cosmology, in which the effective gravitational coupling 
$G_{\rm eff}$ depends  on time, has been the subject of much work 
\cite{FujiiMaeda03,  mybook} but much less attention has been 
devoted to inhomogeneous  solutions in which $G_{\rm eff}$ depends 
also on space. However,  there is really no support for assuming 
that this spatial  dependence can be neglected 
\cite{BarrowOToole01,  CliftonMotaBarrow05}. Spherically 
symmetric inhomogeneous  solutions of scalar-tensor gravity 
representing a central condensation embedded in a cosmological 
background have been  found in \cite{CliftonMotaBarrow05}.

There is plenty of additional motivation for studying analytical 
solutions of gravitational theories representing a central object 
in a cosmological space. First, the present acceleration of the 
cosmic expansion \cite{SN} requires, if one is to  remain within 
the boundaries of general relativity,  that 
approximately 73\% of  the energy 
content of the universe is in the form of exotic 
(pressure $P^{(m)}  \sim -\rho^{(m)}$) dark energy 
\cite{Komatsu} 
(see \cite{LinderResourceLetter} 
for 
a list of references and \cite{AmendolaTsujikawabook} for a  
comprehensive discussion). An alternative to this  {\em ad hoc} 
explanation is that gravity deviates from general  relativity at 
large scales. 
Further motivation for alternative gravity comes from the fact 
that virtually all theories attempting to quantize gravity 
produce, in the  low-energy limit,  not general relativity but 
modifications of it  containing corrections such as non-minimally 
coupled dilatons  and/or higher 
derivative terms. 

These ideas have led to the introduction 
(or better, revival) of $f(R)$ gravity to replace Einstein 
theory at large scales \cite{CCT, CDDT, Vollick, 
metricaffine} and explain the cosmic 
acceleration (see  \cite{review, DeFeliceTsujikawa} for reviews 
and  \cite{otherreviews} for shorter introductions). Since the 
$f(R)$  theories of interest for cosmology are designed to 
produce a  time-varying effective cosmological ``constant'', 
spherically  symmetric solutions representing black holes or 
central condensations in these theories are expected to be 
asymptotically Friedmann-Lema\^itre-Robertson-Walker 
(FLRW), not asymptotically flat, and to be 
dynamical. Very few such  solutions are known, among them the 
inhomogeneous time-dependent  solution of Clifton in  
$f(R)=R^{1+\delta}$ gravity  \cite{Clifton, myClifton}.  

Second, analytical solutions describing central objects in a 
cosmological background are of interest also in general 
relativity. The first study of this kind of solution by 
McVittie \cite{McVittie} is related to investigations of the 
problem of whether, and to what extent,  the cosmic expansion 
affects local systems (see \cite{CarreraGiulini} for a 
recent review). 
In  addition to the old (and largely 
overlooked) McVittie solution \cite{McVittie}, 
which is not yet completely understood 
\cite{Klebanetal, LakeAbdelqader, Roshina, FZN} relatively few 
other solutions with similar features have been reported 
over the years \cite{cosmologicalblackholes}. 

Third, more recent interest in cosmological condensations in the 
context of general relativity arises from yet another attempt to 
explain the present cosmic acceleration without dark energy and 
without modifying gravity. This is the idea that the backreaction 
of inhomogeneities on the cosmic dynamics is sufficient to 
produce the observed acceleration \cite{backreaction}. However, 
the formalism implementing this idea is plagued by formal 
problems and it has not been 
shown  to be able to explain convincingly the cosmic 
acceleration. Indeed, even the sign of the backreaction terms in the 
equation giving the averaged acceleration has not been shown to 
be the correct one  \cite{thatreview, Larenagauge,  
VitaglianoLiberatiFaraoni09} and recent work casts even more serious 
doubts on this proposed solution to the cosmic acceleration problem 
\cite{GreenWald}.   

Attempts to move beyond these riddles involve the 
consideration of analytical solutions of Einstein theory 
describing cosmological inhomogeneities and including 
Lema\^itre-Tolman-Bondi, Swiss-cheese, and other models 
\cite{backreactionexact, Krasinskibook}. Moreover, the 
teleological nature of the event horizon has prompted the 
consideration of apparent, trapping, isolated, dynamical, 
and slowly evolving horizons (\cite{dynamicalhorizons, 
Nielsenreview} and references therein), a subject of great 
interest \cite{dynamicalBHthermo}. There has also been 
interest in dynamical black hole horizons in relation to 
the accretion of dark energy \cite{accretion}.  
Physically, black hole event horizons can only be traversed 
from the outside to the inside while, for the traditional 
cosmological event and particle horizons, signals can cross 
from the inside to the outside but not {\em vice-versa}. 
Event horizons are null surfaces and are appropriate to 
describe stationary situations but, as said, they require 
the knowledge of the entire spacetime manifold to even be 
defined. An apparent horizon is a spacelike or timelike 
surface defined as the closure of a 3-surface which is 
foliated by marginal surfaces (those on which the expansion 
of the congruence of radial null geodesics vanishes) 
\cite{HaywardPRD49}.

With all these motivations in mind, it is 
interesting to further explore  analytical solutions of 
alternative gravity theories representing spherical objects  in 
cosmological  backgrounds. Here we consider the class of 
solutions discovered  by Clifton, Mota, and Barrow 
\cite{CliftonMotaBarrow05} in  Brans--Dicke theory,  described by 
the action \cite{BransDicke61}
\be
S_{BD}=\int d^4x \, \sqrt{-g} \left[ \phi R -\frac{\omega_0}{\phi} 
\, g^{\mu\nu} \nabla_{\mu}\phi \nabla_{\nu}\phi +2\kappa \,{\cal 
L}^{(m)} \right] \,,
\ee
where $\kappa \equiv 8\pi G$, $G$ is Newton's constant, ${\cal 
L}^{(m)}$ is the matter Lagrangian, and the Brans--Dicke scalar 
field $\phi$ corresponds to the inverse of the gravitational 
coupling $G_{\rm eff}$.\footnote{We follow the notations and 
conventions of Ref.~\cite{Wald}.} Matter is assumed to be a 
perfect fluid with 
energy density $\rho^{(m)}$, pressure $P^{(m)}$, and equation of 
state $P^{(m)}=\left( \gamma -1 \right) \rho^{(m)}$, where 
$\gamma$ is a constant \cite{CliftonMotaBarrow05}. In the 
following sections we analyze and discuss the structure of the 
solutions of \cite{CliftonMotaBarrow05}, focussing on the 
dynamical behaviour of their apparent horizons, in an attempt to understand if the these solution harbor black holes or naked singularities. The  bizarre  
behaviour  of the apparent horizons we find seems to be rather typical  of solutions 
describing cosmological black holes \cite{HusainMartinezNunez} in a 
certain region of the parameter space, but other behaviours appear for 
different combinations of the parameters.

Note that, even though it is standard 
procedure to rely on apparent horizons as proxies for event horizons to 
characterize black holes in theoretical and numerical relativity 
\cite{Nielsenreview, proxies}, it is also 
well known that apparent horizons depend on the spacetime slicing 
adopted \cite{foliationdependence} (this problem is perhaps less worrisome  
when spherical symmetry is assumed).   We adopt the same 
practice here, bearing  the {\em 
caveat} just mentioned in mind.

\section{Clifton-Mota-Barrow solutions}

We begin with the Clifton-Mota-Barrow spherically symmetric 
and time-dependent metric \cite{CliftonMotaBarrow05}
\be\label{CMBmetric}
ds^2=-e^{\nu (\varrho)}dt^2+a^2(t) e^{\mu (\varrho)}(d\varrho^2+\varrho^2d\Omega^2) \,,
\ee
where $d\Omega^2= d\theta^2 +\sin^2 \theta \, d\varphi^2 $ denotes 
the line element 
on the unit 2-sphere, 
\bea 
e^{\nu (\varrho)} & = &  
\l(\frac{1-\frac{m}{2\alpha \varrho}}{1+\frac{m}{2 \alpha \varrho}}\r)^{2\alpha 
}\equiv A^{2\alpha} \,,\\
&&\nonumber\\
e^{\mu (\varrho)} & = & \l(1+\frac{m}{2\alpha \varrho}\r)^{4} 
A^{\frac{2}{\alpha}( \alpha-1)(\alpha +2)} \,, \\
&&\nonumber\\
\label{abeta}
a(t) & = & a_0\l(\frac{t}{t_0}\r)^{\frac{ 
2\omega_0(2-\gamma)+2}{3\omega_0\gamma(2-\gamma)+4}}\equiv 
a_{\ast}t^{\beta} \,,\\
&&\nonumber\\
\label{scalart}
\phi(t, \varrho) &= & \phi_0\l(\frac{t}{t_0}\r)^{\frac{2(4-3\gamma)}{ 
3\omega_0\gamma(2-\gamma)+4}}A^{-\frac{2}{\alpha }(\alpha^2-1)} \,,\\
&&\nonumber\\
\alpha & = & \sqrt{ \frac{ 2( \omega_0+2 )}{2\omega_0 +3} } 
\,,\label{7}\\
&&\nonumber\\
\rho^{(m)}(t, \varrho) & = & \rho_0^{(m)} \left( \frac{ a_0}{a(t)} 
\right)^{3\gamma} A^{-2\alpha} \,, \label{density}
\end{eqnarray}
$\rho^{(m)}$ is the energy density of the cosmic fluid, $\omega_0$ 
is the Brans--Dicke parameter, $m$ is a mass parameter, $\alpha, 
\phi_0, a_0$, $\rho^{(m)}_0$ and 
$t_0$ are positive constants (where $\phi_0$, $\rho^{(m)}_0$ and 
$t_0$ are not actually fully independent).   Moreover, $\varrho$ is the isotropic 
radius  related to the  Schwarzschild 
radial coordinate $\tilde{r}$ by 
\be\label{isotropicradius}
\tilde{r} \equiv \varrho\l(1+\frac{m}{2\alpha \varrho}\r)^{2}\,,
\ee
so that
\be
d\tilde{r}=\l(1-\frac{m^2}{4\alpha^2\varrho^2}\r)d\varrho \,.
\ee

The quantity $\alpha $ is real for $\omega_0 <-2$ and for $ \omega_0> 
-3/2$. For definiteness, we impose that $\omega_0>-3/2$ and $ \beta \geq 
0$. 
The Clifton-Mota-Barrow metric~(\ref{CMBmetric}) is separable and 
reduces to the spatially flat Friedmann--Lema\^itre--Robertson--Walker (FLRW)  metric in the limit 
$m\rightarrow 0$ in which the central 
mass disappears.  For $\gamma \neq 2$, setting 
$\omega_0=\left( \gamma -2 \right)^{-1}$ yields $\beta =0$ and 
the metric becomes static, whereas the scalar field remains time-dependent. 
 Setting $\gamma=2$ or $\gamma=4/3$ leads to $\beta =1/2$ and the 
scale factor scales as $\sqrt{t}$ independent of the value of the 
Brans--Dicke coupling  $\omega_0$. We will consider the physically interesting special cases in a separate section below.

Our main concern here is whether the solutions in 
the Clifton--Mota--Barrow class represent black holes or 
naked singularities, embedded in a 
cosmological background (by naked singularity we simply mean a 
timelike or null singularity  which is not covered by an 
event horizon). 
To answer this question, we would like to determine the location and the nature of horizons. Since the spacetime  is 
dynamical, it is appropriate to consider apparent, instead of event, 
horizons and, therefore, we will locate the apparent horizons and study 
their dynamics.

\section{Finding the apparent horizons}

We proceed by rewriting the metric in the more familiar form
\be
ds^2=-A^{2\alpha}dt^2+a^2(t) 
A^{\frac{2}{\alpha}(\alpha^2-2)}d\tilde{r}^2 
+r^2 d\Omega^2 \,,
\ee
using the areal radius  
\bea\label{arealradius}
r & = &  a (t) \varrho \l( 1+\frac{m}{2\alpha \varrho} \r)^{2} 
A^{\frac{1}{\alpha}(\alpha -1)(\alpha+2)} \nonumber\\
&&\nonumber\\
& = &  a(t) \tilde{r} 
A^{\frac{1}{\alpha}( \alpha -1)( \alpha +2)} \,.
\eea

The differential $dr$ is related to  $d\tilde{r}$ by 
\begin{eqnarray}
d r & = &  \dot{a}(t) \,  \tilde{r} A^{\frac{1}{\alpha}( \alpha 
-1)( \alpha +2)} dt 
+ a (t) A^{\frac{1}{\alpha}( \alpha-1)( \alpha +2)} 
d\tilde{r} 
\nonumber\\
&&\nonumber\\
&  &+ 
\frac{a(t) m}{\alpha^2\tilde{r}}( \alpha -1)( \alpha 
+2)A^{\frac{1}{\alpha}( \alpha -1)( \alpha +2)-2} 
d\tilde{r} \,,\,\,\,
\end{eqnarray}
which means that 
\be
d\tilde{r}=\frac{dr-\dot{a}(t)\tilde{r} 
A^{\frac{1}{\alpha}( \alpha -1)( \alpha +2)}dt}{a (t)
A^{\frac{1}{\alpha}( \alpha -1)( \alpha +2)-2} 
\l[A^2+\frac{m}{\alpha^2\tilde{r}}( \alpha -1)( \alpha +2)\r]} \,. 
\ee

The line element can now be written as 
\begin{eqnarray}
ds^2 &=& -\l[A^{2\alpha}-\frac{\dot{a}^2(t)\tilde{r}^2}{B^2(\varrho)} 
A^{\frac{2}{\alpha}( \alpha^2+2\alpha -2)}\r] dt^2 
\nonumber\\
&&\nonumber\\
& \, & -  2 \, \frac{\dot{a}(t)\tilde{r}}{B^2(\varrho)} 
A^{\frac{\alpha^2+3\alpha -2}{\alpha}}dr dt \nonumber\\
&&\nonumber\\
&\, & +\frac{A^2(\varrho)}{B^2(\varrho)}dr^2 
+r^2d\Omega^2 \,,
\end{eqnarray}
where we have defined the positive 
function
\be \label{B(r)}
B(\varrho) \equiv A^2(\varrho)+ \frac{( \alpha -1)( \alpha +2)}{\alpha^2} \, 
\frac{m}{\tilde{r}} \,.
\ee
(Note that $B>0$ is a 
consequence of $\alpha=\sqrt{\frac{2(\omega_0 + 
2)}{2\omega_0 +3 }}\geq1$.)

We now introduce a new time coordinate  $\bar{t}$ which serves 
the purpose of eliminating the time-radius cross term. Define 
$\bar{t}$ such that
\be
d\bar{t}=\frac{1}{F(t, r )}\left[ dt+\psi(t, r )dr 
\right]  \,,
\ee
where $\psi (t, r) $ is a function to be fixed later and  $F(t, 
r )$ is an integrating factor which guarantees 
that $d\bar{t}$ is an exact differential and satisfying the 
equation
\be
\frac{ \partial }{ \partial r}\left( \frac{1}{F} \right)=
\frac{ \partial}{ \partial t } \left( \frac{\psi}{F} \right) \,.
\ee

The line element then assumes the form
\begin{widetext}
\bea
ds^2 & = & -\l[A^{2\alpha}-\frac{\dot{a}^2(t)\tilde{r}^2}{B(\varrho)^2} 
A^{\frac{2}{\alpha}(\alpha^2+2\alpha -2)}\r] F^2d\bar{t}^2 
+  \l\{2\psi 
F\l[A^{2\alpha}-\frac{\dot{a}^2(t) \tilde{r}^2}{B(\varrho)^2} 
A^{\frac{2}{\alpha}(\alpha^2+2\alpha -2)}\r]-2 \, 
\frac{F\dot{a}(t) \tilde{r}}{B(\varrho)^2} 
A^{\frac{\alpha^2+3\alpha-2}{\alpha}}\r\}dr d\bar{t}
\nonumber\\&&\nonumber\\&\, & 
+ \l\{\frac{A^2}{B(\varrho)^2}-\psi^2\l[A^{2\alpha} 
-\frac{\dot{a}^2(t) 
\tilde{r}^2}{B(\varrho)^2}A^{\frac{2}{\alpha}(\alpha^2+2\alpha 
-2)}\r] 
+2 \, \frac{\psi \dot{a}(t)\tilde{r}}{B(\varrho)^2} 
A^{\frac{\alpha^2+3\alpha -2}{\alpha}}\r\}dr^2+r^2 d\Omega^2 
\,.
\eea
\end{widetext}

The choice
\be
\psi=\frac{\dot{a}(t) \tilde{r}}{B^2} 
\frac{A^{\frac{-\alpha^2+3\alpha -2}{\alpha}}}{D(t,\varrho)}
\ee
for the function $\psi$, with 
\be
D(t, \varrho)\equiv 1-\frac{\dot{a}^2(t) 
\tilde{r}^2}{B^2}A^{\frac{4}{\alpha}( 
\alpha -1)}\,,
\ee
turns the metric into the simple form
\begin{eqnarray}
ds^2 & = & -A^{2\alpha}DF^2d\bar{t}^2+\l(\frac{H^2}{B^4D}r^2 
A^{2(2-\alpha )}+\frac{A^2}{B^2}\r)dr^2 \nonumber\\
&&\nonumber\\
& &+ r^2d\Omega^2  \,,
\end{eqnarray}
where $H \equiv \dot{a}(t)/a(t) $ denotes the Hubble parameter of the 
background FLRW  universe. We are now able to locate the apparent 
horizons (when they exist), which are the loci of spacetime 
points satisfying $\nabla^c r \nabla_c r =0$, or  $g^{r 
r}=0$ \cite{NielsenVisser, Nielsenreview}, that is 
\be
\frac{B^4D}{H^2 r^2 A^{2(2-\alpha)}+A^2B^2D}=0 \,.
\ee

The solution of this equation reduces to the condition $D=0$, or
\be
\label{bahr}
B^2A^{2(\alpha-1)}=H^2 r^2\,,
\ee
which, explicitly, reads 
\be\label{ThisOne}
A^{\alpha-1}\!\l[A^2\!+\frac{(\alpha -1)( \alpha +2)}{\alpha^2}  \, 
\frac{m a(t)}{r} \, A^{\frac{(\alpha -1)(\alpha +2)}{\alpha }}\r]\!=
\pm Hr  \,.
\ee

In an expanding universe with $H>0$ the quantity in square  brackets  is positive, hence we choose the positive sign.  
Eq.~(\ref{ThisOne}) can then be written as 
\be \label{19}
Hr^2-\frac{(\alpha -1)(\alpha +2)}{\alpha^2} \, m \, a(t) 
A^{\frac{2(\alpha -1)( \alpha +1)}{\alpha}}-A^{\alpha +1}r =0 \,.
\ee

The Ricci scalar becomes  singular as $r \rightarrow 0$ 
for all positive values of the mass parameter $m$ (see 
Appendix~A) and this limit denotes a central singularity.  The 
energy density~(\ref{density}) of the cosmic fluid also diverges 
in this limit.

\section{Special cases and limits}

Before determining the generic behaviour of apparent horizons, it is useful to look into some special limits, of either the theory or the solutions, which will help us gain some intuition.

\subsection{The zero mass limit}

In the limit $m\rightarrow 0 $ 
in which there is no central object, eq.~(\ref{19}) reduces to 
$H r^2=r$, which yields $r =H^{-1}$, the Hubble 
horizon.\footnote{In a FLRW universe with curvature index 
$k\neq 0$ the cosmological apparent horizon has radius $\left( 
H^2+k/a^2 \right)^{-1/2}$.} 
This value is also obtained in the limit of large $\varrho$ in which  
$r$ becomes  a comoving radius and  the metric approaches the 
spatially flat FLRW metric. This is best seen using eq.~(\ref{bahr}) as at this limit $A,B\to 1$ (the limit is less straightforward in eq.~(\ref{19}) as $r\to \infty$ and $\varrho\to \infty$).
Therefore, we expect the horizon at larger radii to be a cosmological one.

\subsection{The static limit}

We now consider the limit in which the metric becomes static, which corresponds to $\beta=0$ and yields $ 
a(t)\equiv a_0 $, see eq.~(\ref{abeta}). This value for $\beta$ is obtained for 
$\omega_0=\left( \gamma -2\right)^{-1}$ (with $\gamma\neq 
2$). This requirement implies that for each theory in the 
Brans--Dicke class, ({\em i.e.}, for each value of 
$\omega_0$) there is at most one solution with a static 
metric in the Clifton--Mota--Barrow family, and it 
corresponds to a specific choice of equation of state for 
the fluid. As mentioned earlier, for $\alpha$ to be real, 
one needs to have $\omega_0<-2$ or $\omega_0 >-3/2$.  This  
translates to $\gamma >3/2$ or $\gamma<4/3$ when the 
$\beta=0$ condition has been imposed.  

Eqs.~(\ref{scalart}) and (\ref{density}) yield
\begin{eqnarray}
 \phi( t, r) &=& \phi_0 \left( \frac{t}{t_0} \right)^{2} 
A^{-\frac{2( \alpha^2 -1)}{\alpha} } \,,\label{static2}\\
&&\nonumber\\
\rho^{(m)}&=& \rho_0^{(m)}  A^{-2\alpha} \,. \label{static3}
\end{eqnarray}

The Brans--Dicke field $\phi$ depends on time even though the 
metric $g_{\mu\nu}$ and the matter energy density 
$\rho^{(m)}$ do not. In fact, there is no solution in 
the Clifton--Mota--Barrow class which is genuinely static.

  In terms of the 
areal radius $r$, it is
\be 
ds^2=-A^{2\alpha}dt^2 +\frac{A^2}{B^2} dr^2 +r^2 
d\Omega^2 \,,
\ee
and the apparent horizons are located by the equation $g^{r 
r}=0 $ equivalent to $B=0$, or 
\be
\varrho^2 +\frac{m}{\alpha^2} \left( \alpha^2-2 \right)\varrho 
+\frac{m^2}{4\alpha^2}=0 \,.
\ee

The discriminant of this quadratic equation is $\Delta (\alpha^2) = 
\frac{m^2}{\alpha^2} \left[ \left( \alpha^2-2\right)^2 -\alpha^2 
\right]$ and one easily finds that $\Delta \geq 0$ for $\alpha 
\leq 1 $ and $ \alpha \geq 2$ (remember that $\alpha \geq 0$, cf. 
eq.~(\ref{7})). 
Therefore, for $ 1<\alpha < 2$ 
there are no real roots and no apparent horizons. For $ \alpha \leq 
1$ and for $ \alpha \geq 2$ the real roots, 
\be
\varrho_{\pm}= \frac{m}{\alpha^2} \left[ -\left( \alpha^2 -2 
\right)\pm \sqrt{ \left( \alpha^2 -2 \right)^2 -\alpha^2 } 
\right]\,,
\ee
are both negative and do not correspond to apparent horizons. We 
conclude that the solution with static  metric always describes a 
naked singularity. 

\subsection{The  limit to general relativity}
\label{grlimit}

We now consider the limit to general relativity obtained for $\omega_0 \rightarrow 
\infty$. When $\gamma \neq 0$ and $\gamma\neq 2$, this 
limit yields $\alpha \rightarrow 1$, $\phi \rightarrow \phi_0$,  
and\footnote{When $\gamma=2$ the scale 
factor  is forced to be $a(t)\propto \sqrt{t} $, $\phi\propto t^{-1}$ 
and $\rho^{(m)}\propto t^{-3}$ and do not depend 
on the Brans--Dicke coupling parameter $\omega_0$. However, the limit 
$\omega_0\rightarrow \infty$ still yields $\alpha=1$ and 
leads to the same functional dependence  on $\varrho$ for the 
various quantities as the $\gamma\neq 2$ case. The metric 
still belongs to the generalized McVittie class.} 
\begin{eqnarray} 
\label{grlm}
ds^2 & = & - \left( \frac{ 1-\frac{m}{2\varrho} }{ 1+\frac{m}{2\varrho} 
}\right)^2 dt^2 +  
a^2(t) \left( 1+\frac{m}{2\varrho} \right)^4 \cdot \nonumber\\
&&\nonumber\\
& {\left. \right.}&  \cdot \left( 
d\varrho^2+\varrho^2 d\Omega^2 \right) \,,\\
&&\nonumber\\
a(t) & = & a_0 \left( \frac{t}{t_0} \right)^{ \frac{2}{3\gamma}} 
\,,\\
&& \nonumber\\
\rho^{(m)}(t) & = & \rho_0^{(m)} \left( \frac{t_0}{t} \right)^2 
A^{-2}\,.
\end{eqnarray}

This metric corresponds to one of the generalized McVittie metrics 
studied in  Refs.~\cite{generalizedMcVittie, 
generalizedMcVittie2, McVittieattractor} 
which, in isotropic coordinates, assume the form
\begin{eqnarray} 
ds^2 &= &  -\left( \frac{  1-\frac{M(t)}{2\varrho a(t)} }{ 
1+\frac{M(t)}{2\varrho a(t)} 
}\right)^2  dt^2 \nonumber\\
&&\nonumber\\
& \left.\right. & +  a^2(t) \left( 1+\frac{M(t)}{2 \varrho a(t)} 
\right)^4 \left( d\varrho^2+\varrho^2 d\Omega^2 \right) \,,
\label{generalizedMcVittie}
\end{eqnarray}
where $M(t)$ is an arbitrary positive regular function of time. 

The  
McVittie solution of general relativity originally introduced to 
study the effect of the cosmological expansion on local systems 
\cite{McVittie}  is obtained for  $ M(t)= $~const. This 
time independence of the  function $M(t)$ in this case follows from the 
McVittie condition $G^1_0=0$ which corresponds to zero radial 
energy flow $T_0^1=0$. Lifting this restriction and allowing for 
radial accretion of energy  generates the 
solutions~(\ref{generalizedMcVittie}) with general functions $M(t)$  
(see the discussion in Ref.~\cite{generalizedMcVittie}). It is 
shown in Ref.~\cite{McVittieattractor} that, in the 
class of generalized McVittie  solutions~(\ref{generalizedMcVittie}), the 
solution with ``comoving mass function'' $M(t)=M_0 a(t)$ (where  
$M_0$ is a constant) is a late-time  attractor for solutions 
characterized by a background 
universe which keeps expanding in the 
future ($a\rightarrow  +\infty$). This is  precisely the 
$\omega_0\rightarrow  \infty$ limit of the scalar-tensor 
solution~(\ref{CMBmetric})-(\ref{density}), which also 
makes it clear that the Clifton--Mota--Barrow solutions 
are indeed accreting. Incidentally, the generalized 
McVittie 
solutions~(\ref{generalizedMcVittie}) of general relativity  were  
derived two  years after the discovery of the Clifton-Mota-Barrow 
solution~(\ref{CMBmetric})-(\ref{density}) and the one with late-time 
attractor behaviour and with $M=M_0 \, a(t)$ could, in principle, 
have been  discovered by taking the limit to general relativity of 
this Brans--Dicke solution. 

\begin{figure}[t]
\includegraphics[scale=0.88]{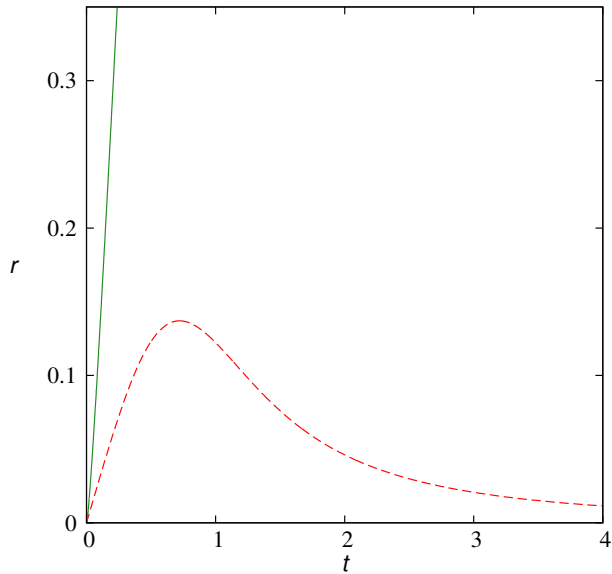}
\caption{\label{fig:w1712}
 Radii of the apparent horizons in units of $(m a_*)^{1/(1-\beta)}$ as 
functions of time in the same units for $\omega_0=-17/12$. The 
red, dashed curve corresponds to dust ($\gamma=1$) and the green, solid curve corresponds to both radiation ($\gamma=4/3$) and stiff matter  
($\gamma= 2$).  For dust,  there is only one 
apparent horizon whose radius reaches a maximum and then decreases. For radiation and stiff matter, instead, there is a naked singularity in a  
universe which expands forever.}
\end{figure}

The apparent horizons of the generalized McVittie 
metrics~(\ref{generalizedMcVittie}) have been discussed in 
\cite{generalizedMcVittie2}. For large values of $\omega_0$, the 
solution~(\ref{CMBmetric})-(\ref{density}) approaches the attractor 
McVittie solution and its apparent horizons should also approach 
those of the attractor McVittie metric: jumping ahead slightly, this is indeed the case, as 
can be seen by comparing our Fig.~\ref{fig:winf} with Fig.~3 of 
Ref.~\cite{generalizedMcVittie2}.

The $\gamma=0$ case, which corresponds to a cosmological constant, 
leads to a diverging exponent $\beta$ for the scale factor $a(t)$ when 
$\omega_0 \to \infty$. This behaviour can be attributed to the fact 
that the Clifton--Mota--Barrow solution assumes a power law form for 
the scale factor, whereas the general relativity limit of the solution 
is actually expected to be Schwarzschild--de Sitter spacetime.

For $\gamma=2$ the $\omega_0\to \infty$ limit yields $\alpha\to 1$, 
$\phi\propto t^{-1}$, $a(t)\propto \sqrt{t}$, $\rho^{(m)}(t)\propto t^{-3} A^{-2}$, and the metric is the same as in eq.~(\ref{grlm}).

\begin{figure}[t]
\includegraphics[scale=0.88]{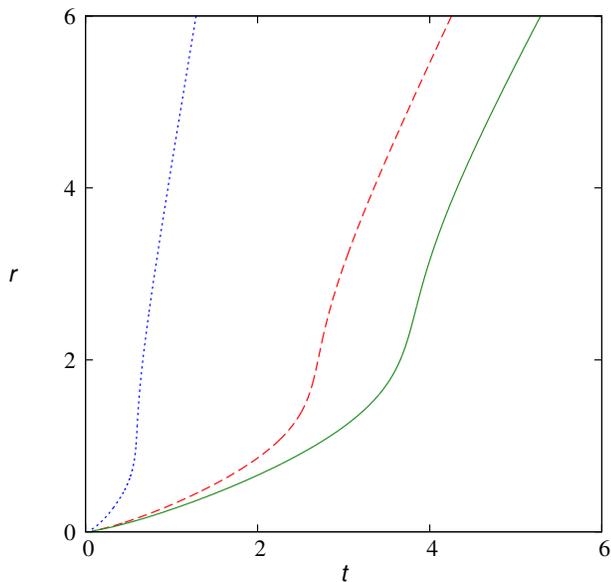}
\caption{Same as Fig.~\ref{fig:w1712} but for $\omega_0=-1/3$. The blue, dotted line corresponds to a cosmological constant ($\gamma=0$). In all cases there is one, ever expanding horizon, and so the solution appears to represents a naked singularity in an expanding universe.}
\label{fig:w13}
\end{figure}

\section{Generic behaviour of apparent horizons}

Having discussed the special cases, we now turn our 
attention to the behaviour of apparent horizons in 
generic solutions of the Clifton--Mota--Barrow family. In 
order to solve  eq.~(\ref{19}) and determine the location of these horizons, it is convenient to 
introduce the new quantity  $x\equiv \frac{m}{2\alpha \varrho}$, in terms 
of which it is 
\be
A=\frac{1-x}{1+x} \,,
\ee
while $  H=\beta / t $. One can now express parametrically the radius 
$r$ of the apparent horizon(s) and the time coordinate  as 
functions of the parameter $x$, obtaining
\begin{eqnarray}
r (x) & = & a_{\ast} t^\beta \frac{m}{2\alpha} 
\, \frac{(1+x)^2}{x} \l(\frac{1-x}{1+x}\r)^{\frac{(\alpha -1)( \alpha 
+2)}{\alpha}}   \,,\\
&&\nonumber\\
t(x) &  =& \l\{ \frac{2\alpha}{m \, a_{\ast}\beta} \, 
\frac{x}{(1+x)^{\frac{2}{\alpha}(\alpha +1)}}\l[
(1-x)^{2/\alpha} \right.\right. \nonumber\\
&&\nonumber\\
&+ & \left. \left.  2x \, \frac{(\alpha -1)(\alpha +2)}{\alpha}  \, 
(1-x)^{-2(\alpha - 1)/\alpha }\r] \r\}^{\frac{1}{\beta-1}}\,. \nonumber\\
\end{eqnarray}

\begin{figure}[ht]
\centering
\subfigure[~$\omega_0=1$]{
\includegraphics[scale=0.88]{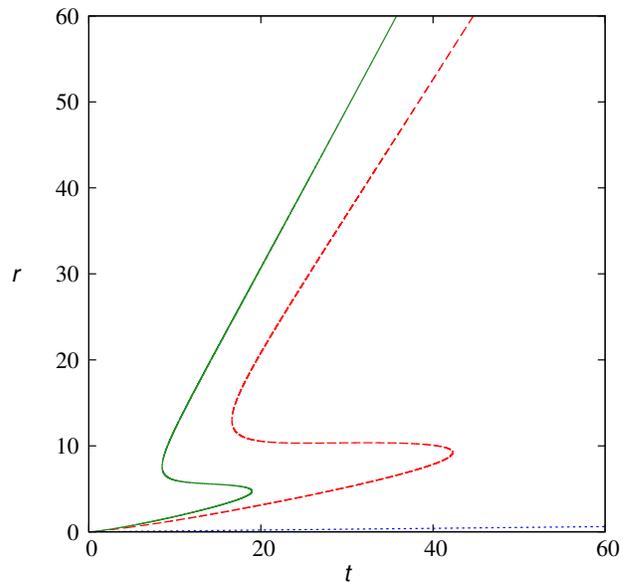}
\label{fig:w1a}
}\\
\subfigure[~$\omega_0=1$, zoom-in]{
\includegraphics[scale=0.88]{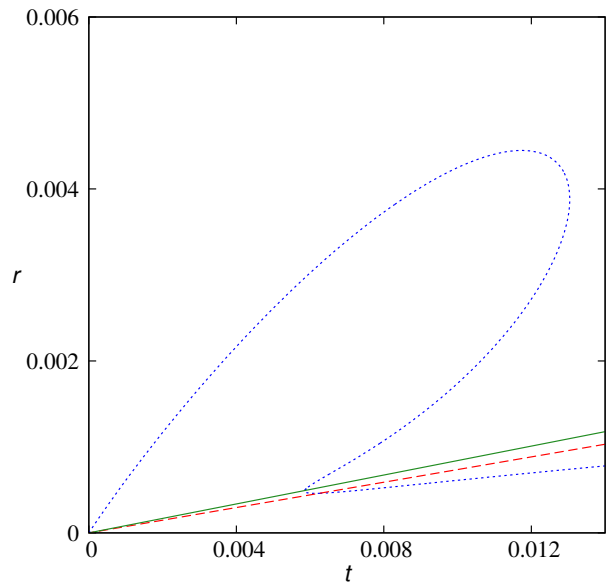}
\label{fig:w1zoom}
}
\caption{Same as previously but for $\omega_0=1$. For all three values 
of $\gamma$, at  early times there  is only one horizon. As the 
universe  expands, the singularity gets covered by two more apparent 
horizons. Two of the horizons eventually merge and disappear, leaving behind only the 
cosmological horizon covering a naked 
singularity.}
\label{fig:w1}
\end{figure}

\begin{figure}[t]
\centering
\subfigure[~$\omega_0=10^5$]{
\includegraphics[scale=0.88]{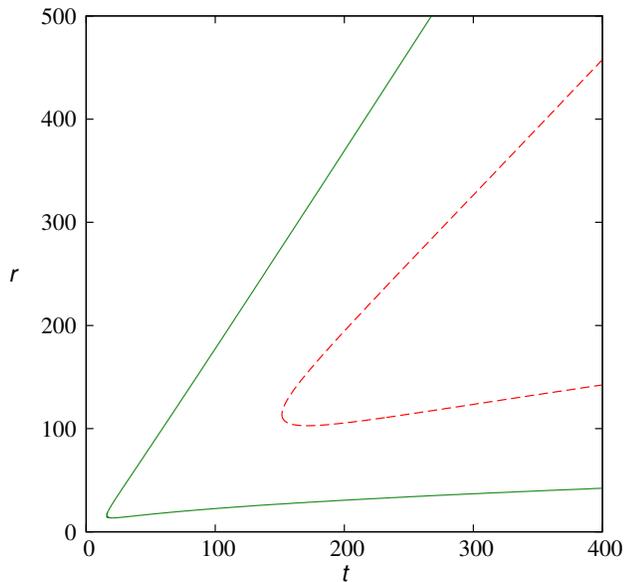}
\label{fig:winf1}
}\\
\subfigure[~$\omega_0=10^5$, zoom-in]{
\includegraphics[scale=0.88]{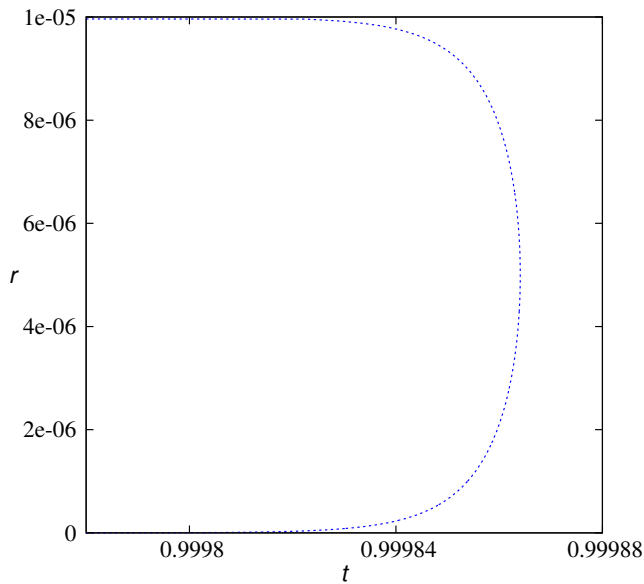}
\label{fig:winfzoom}
}
\caption{Same as previously but for $\omega_0=10^5$, $\omega_0\to \infty$ being the limit to general relativity. Here for all three cases there are two horizons, presumably a black hole horizon and a cosmological horizon. For a cosmological constant (blue, dotted curve) these two horizon merge and disappear. For the other two cases, there is initially a naked singularity which eventually gets covered by the two horizons.}
\label{fig:winf}
\end{figure}

The radii of the apparent horizons as functions of time are plotted in Figs.~\ref{fig:w1712} to \ref{fig:winf}  for the values of the Brans--Dicke 
parameter $\omega_0= -17/12$, $-1/3$, $1$, and $10^5$, 
respectively, and for various choices of the equation of 
state parameter $\gamma$. In these plots $r$ and $t$ are actually measured in units of 
\begin{equation}
(m a_*)^\frac{1}{1-\beta}=\left(a_0\, 
\frac{m}{t_0}\right)^\frac{1}{1-\beta} t_0\,,
\end{equation}
as this convenient normalization completely absorbs the dependence on the parameters $m$, $a_0$, $t_0$.

 The  
blue, dotted  curves correspond to a cosmological  
constant ($\gamma=0$) and the 
red, dashed curves correspond to dust ($\gamma=1$). The green, solid curves show the 
behaviour of the apparent horizons for both radiation ($\gamma=4/3$) and stiff matter  
($\gamma= 2$). This is because $\beta$, which determines the scaling of the scale factor with time,  is equal to $1/2$ and independent of $\omega_0$ for both of these values. For $\omega_0=-17/12$ (Fig.~\ref{fig:w1712}) we do not consider the case of a cosmological constant, corresponding to $\gamma=0$, as it leads to a contracting universe.

As can be seen in the figures (see captions for more details), for $\omega_0=-17/12$ and $\omega_0=-1/3$ there is only one apparent horizon for all of the values of $\gamma$ we have considered. In most cases, this horizon is expanding forever, so the solution is most likely to represent a naked singularity in an expanding universe. For $\omega_0=-17/12$ and for dust ($\gamma=1$), on the other hand, the apparent horizon exhibits a perhaps more remarkable behaviour: it initially expands, to reach a maximum radius and then contracts to reach zero radius asymptotically. 

Even more noteworthy is the behaviour of the apparent horizons when  
$\omega_0=1$ (Fig.~\ref{fig:w1}). For dust, radiation, and stiff 
matter  there is initially one expanding apparent horizon, see 
Fig.~\ref{fig:w1a}. Two more apparent horizons appear. The outer 
one expands, while the inner one eventually merges with the initial one 
and they both disappear. Similar 
phenomenology was reported in Ref.~\cite{myClifton} for Clifton's 
solution  \cite{Clifton} of 
metric $f(R)=R^{1+\delta}$ gravity.\footnote{This fact is not 
surprising since metric $f(R)$ gravity 
is equivalent to a Brans--Dicke theory with 
$\phi=f'(R)$, $\omega=0$ and a scalar field potential $V(\phi)$ 
\cite{review}.} 

In fact, this puzzling behaviour 
was found long  ago in the  Husain-Martinez-Nu\~{n}ez solution 
\cite{HusainMartinezNunez} describing a black hole embedded in a 
universe filled with a free massless scalar field minimally 
coupled to gravity and accreting onto the black hole (compare 
Fig.~\ref{fig:w1a} with Fig.~1 of Ref.~\cite{HusainMartinezNunez}). 

For $\omega_0=1$ and $\gamma=0$, which corresponds to a cosmological constant and is presented in Fig.~\ref{fig:w1zoom}, the situation is similar, except for the fact that the pair of horizons actually appears inside the initial horizon. Such behaviour has not been reported before to the best of our knowledge.

 Finally, Fig.~\ref{fig:winf} corresponds to the large 
value of the Brans--Dicke parameter $\omega_0=10^5$. The behaviour of the  
apparent horizon dynamics  is very similar to that present in  the 
general relativity limit of the Clifton-Mota-Barrow solution 
obtained for $\omega_0 \rightarrow \infty$ and discussed in 
Sec.~\ref{grlimit}.
For dust, radiation and stiff matter, the singularity is initially naked and eventually gets covered by two expanding horizons, see Fig.~\ref{fig:winf1}. For a cosmological constant this picture is reversed: there are initially two nested horizons, one expanding and one contracting, which eventually merge and disappear, leaving the singularity naked, see Fig.~\ref{fig:winfzoom}.

\section{Discussion and conclusions}

There are relatively few solutions describing central matter 
configurations embedded in FLRW backgrounds in general relativity, and 
even fewer in alternative theories of gravity. We have studied here 
the Clifton--Mota--Barrow class of spacetimes, which are solutions of 
Brans--Dicke theory. The latter  is perhaps the minimal implementation of a varying gravitational coupling, containing only a scalar extra  degree of freedom. As such, it is justly regarded as the prototypical  alternative to Einstein's theory. It is, therefore, quite interesting to assess whether or under which conditions can the Clifton--Mota--Barrow spacetimes describe a realistic localized matter configuration embedded in an evolving universe.

Given that these spacetimes contain singularities, we have focussed 
our study on the behaviour of dynamical apparent horizons.  According 
to the position in parameter space, we have uncovered different types 
of behaviour for these horizons. The most important result is perhaps 
that, for certain values of the parameters, the Clifton--Mota--Barrow 
spacetime appears to contain a naked singularity (at least 
as far as  one can tell based on the presence/absence of 
apparent horizons; though unlikely, it 
is possible that the particular slicing of the spacetime 
leads to the absence of an apparent horizon even though 
the singularity is cloaked by an event horizon). In 
some cases, this singularity 
is present from the time of the big bang, thus preventing us from 
obtaining the metric and scalar field as regular developments of Cauchy 
data, and later gets covered by black hole and cosmological horizons. For other values of the parameters, pairs of black hole  and cosmological horizons appear and bifurcate, or merge  and disappear, a phenomenology known from a solution of general relativity \cite{HusainMartinezNunez} and one of $f(R)$ gravity \cite{Clifton, myClifton}. Overall, the Clifton--Mota--Barrow class of solutions exhibits a great richness of behaviours of its apparent horizons, including 
the new ones reported in Fig.~\ref{fig:w1712} and Fig.~\ref{fig:w1}.

The physical relevance of spacetimes harbouring naked singularities is, 
of course, questionable. However, there are still two scenarios in 
which the Clifton--Mota--Barrow spacetimes might still be physically 
relevant: (i) in the region of the parameter space where a black hole 
horizon eventually cloaks the singularity, it is conceivable that they 
can (approximately) describe the late time evolution of black holes 
that have formed from collapse in FLRW spacetime (a different solution 
would be needed to describe this collapse); (ii) even in the region 
where no horizon forms, they might be able to (approximately) describe 
the exterior of a matter configuration embedded in an FLRW universe (a 
different solution will be needed in order to describe the interior). 
Whether or not any of these two scenarios are meaningful requires 
further investigation.

The fact that such a variety of behaviours (cosmological black holes, 
naked singularities, appearing/bifurcating and merging/disappearing 
pairs of apparent horizons) is contained in the relatively simple 
Brans-Dicke theory leads us to believe that more complicated 
theories of gravity will exhibit an even greater degree of richness 
and complication when it comes to dynamical horizons, which has not yet 
been explored.

Lastly, one might be tempted to consider the thermodynamics of these 
dynamical apparent horizons, although its physical meaning 
is still questioned \cite{Alexprivate}. In any case, it 
should be noted that the field 
equations of Brans--Dicke theory can be recast in the form of 
effective Einstein equations $G_{\mu\nu}=8\pi ( T_{\mu\nu} + 
T^{(\phi)}_{\mu\nu})$ in which the Brans--Dicke scalar field plays the 
role of an effective stress-energy component $T^{(\phi)}_{\mu\nu}$. 
The latter can easily violate all of the energy conditions because it 
contains terms linear in the second derivatives of $\phi$ in addition 
to the usual terms quadratic in its first derivatives.

\begin{acknowledgments}

We would like to thank John Barrow for pointing out 
the solutions of Ref.~\cite{CliftonMotaBarrow05} and Timothy Clifton 
for enlightening discussions. VF thanks SISSA for its 
hospitality and  the Natural  Sciences and 
Engineering  Research Council of Canada for 
financial support.  VV is supported by FCT - Portugal through grant SFRH/BPD/77678/2011 and would like to  thank Bishop's University for the hospitality 
during the inception of this work. TPS acknowledges partial financial support provided under a Marie Curie Career Integration Grant and the ``Young SISSA Scientists'
Research Project'' scheme 2011-2012, promoted by the
International School for Advanced Studies (SISSA), Trieste, Italy.
\end{acknowledgments}

\appendix\section{Ricci scalar}

The expression of the Ricci scalar is 
\begin{widetext}
\begin{eqnarray}
R &=& -2\left\{ 18 \alpha \varrho \left( 1-\frac{m}{2\alpha \varrho} \right) 
\dot{a}^2 m^6
+576 \varrho^6 \alpha^6 \left( 1-\frac{m}{2\alpha \varrho} \right)^{ 
\frac{2(\alpha -1)(\alpha +2)}{\alpha} } \dot{a}^2 m
+96 \varrho^5 \alpha^5 \left( 1-\frac{m}{2\alpha \varrho} \right)^{ 
\frac{2(\alpha -1)(\alpha +2)}{\alpha} }\dot{a}^2 m^2  
\right.\nonumber\\
&&\nonumber\\
& - & \left. 240 \varrho^4 \alpha^4 \left( 1-\frac{m}{2\alpha \varrho} \right)^{ 
\frac{2(\alpha -1)(\alpha +2)}{\alpha} }\dot{a}^2 m^3
+8 \varrho^3 \alpha^5 \left( 1-\frac{m}{2\alpha \varrho} \right)^{2\alpha} 
\left[ \frac{ 2(\alpha -1)(\alpha +2)}{\alpha} \right]^2 m^2
+ 32 \varrho^3 m^2 \left( 1-\frac{m}{2\alpha \varrho} \right)^{2\alpha} \alpha^7  
\right. 
\nonumber\\
&&\nonumber\\
&-& \left. 96\left( 1-\frac{m}{2\alpha \varrho} \right)^{2\alpha} m^2 \varrho^3  
\alpha^6
+16 \varrho^2 m^3 \left( 1-\frac{m}{2\alpha \varrho} \right)^{2\alpha} \alpha^6 
+16 \varrho^2 m^3 \alpha^5 \left( 1-\frac{m}{2\alpha \varrho} \right)^{2\alpha} 
\right. 
\nonumber\\
&&\nonumber\\
& +& \left. 16 \varrho^3 m^2 \alpha^6 \left( 1-\frac{m}{2\alpha \varrho} 
\right)^{2\alpha} 
\left[ \frac{2(\alpha -1)(\alpha +2)}{\alpha} \right]
+8  \varrho^2 m^3 \alpha^5 \left( 1-\frac{m}{2\alpha \varrho} \right)^{2\alpha}  
 \left[ \frac{2(\alpha -1)(\alpha +2)}{\alpha} \right] \right. 
\nonumber\\
&&\nonumber\\
&- & \left. 96 \varrho^3 \alpha^5 m^2\left( 1-\frac{m}{2\alpha \varrho} 
\right)^{2\alpha}  
 \left[ \frac{2(\alpha -1)(\alpha +2)}{\alpha} \right]  \right. 
\nonumber\\
&&\nonumber\\
&- & \left. 120 \varrho^3 \alpha^3 m^4 \left( 1-\frac{m}{2\alpha \varrho} 
\right)^{ \frac{2(\alpha -1)(\alpha +2)}{\alpha} } \dot{a}^2
+16 \varrho^2 m^3 \alpha^4 \left( 1-\frac{m}{2\alpha \varrho} \right)^{2\alpha}  
 \left[ \frac{2(\alpha -1)(\alpha +2)}{\alpha} \right] \right. 
\nonumber\\
&&\nonumber\\
& + & \left. 4\varrho^2 \alpha^2 m^3 \left( 1-\frac{m}{2\alpha \varrho} 
\right)^{2\alpha}  
 \left[ \frac{2(\alpha -1)(\alpha +2)}{\alpha} \right]^2
+12 \varrho^2 \alpha^2 m^5 \left( 1-\frac{m}{2\alpha \varrho} \right)^{  
\frac{2(\alpha -1)(\alpha +2)}{\alpha} }  \dot{a}^2 \right. 
\nonumber\\
&&\nonumber\\
& + & \left. 384 \varrho^7 \alpha^7  \left( 1-\frac{m}{2\alpha \varrho}  
\right)^{\frac{2(\alpha -1)(\alpha +2)}{\alpha} }  
a \ddot{a}
+576 m \varrho^6 \alpha^6 \left( 1-\frac{m}{2\alpha \varrho}  
\right)^{\frac{2(\alpha -1)(\alpha +2)}{\alpha} } a \,\ddot{a}\right. 
\nonumber\\
&&\nonumber\\
& + & \left. 96 m^2 \varrho^5 \alpha^5 \left( 1-\frac{m}{2\alpha \varrho}  
\right)^{\frac{2(\alpha -1)(\alpha +2)}{\alpha} } a \,\ddot{a}
-240 m^3 \varrho^4 \alpha^4 \left( 1-\frac{m}{2\alpha \varrho}  
\right)^{\frac{2(\alpha -1)(\alpha +2)}{\alpha} } a \,\ddot{a} \right. 
\nonumber\\
&&\nonumber\\
&-& \left. 120 m^4 \varrho^3 \alpha^3 \left( 1-\frac{m}{2\alpha \varrho}  
\right)^{\frac{2(\alpha -1)(\alpha +2)}{\alpha} }a \,\ddot{a} +12  m^5 \varrho^2 
\alpha^2 \left( 1-\frac{m}{2\alpha \varrho}  
\right)^{\frac{2(\alpha -1)(\alpha +2)}{\alpha} }a \,\ddot{a} \right. 
\nonumber\\
&&\nonumber\\
& + & \left. 18 m^6 \varrho \alpha \left( 1-\frac{m}{2\alpha \varrho}  
\right)^{\frac{2(\alpha -1)(\alpha +2)}{\alpha} }a \,\ddot{a}
+128 \varrho^5 \alpha^7 \left( 1-\frac{m}{2\alpha \varrho}  \right)^{2\alpha} \right. 
\nonumber\\
&&\nonumber\\
& +& \left. 3 m^7 \left( 1-\frac{m}{2\alpha \varrho}  \right)^{ 
\frac{2(\alpha -1)(\alpha +2)}{\alpha} } 
a \,\ddot{a}
+384 \varrho^7 \alpha^7 \left( 1-\frac{m}{2\alpha \varrho}  \right)^{ 
\frac{2(\alpha -1)(\alpha +2)}{\alpha} } \dot{a}^2 \right. 
\nonumber\\
&&\nonumber\\
&-& \left. 64 m\varrho^4 \alpha^6 \left( 1-\frac{m}{2\alpha \varrho}  
\right)^{2\alpha}
+3 m^7 \left( 1-\frac{m}{2\alpha \varrho}  \right)^{ 
\frac{2(\alpha -1)(\alpha +2)}{\alpha} } \dot{a}^2 \right\}  \nonumber\\
&&\nonumber\\
&\cdot& \left\{ 4\alpha^2 \varrho^2 a^2 \left( 4\alpha^2\varrho^2 +4\alpha \varrho m+m^2 
\right) \left( 
8\alpha^3 \varrho^3  +12\alpha^2m \varrho^2 +6\alpha m^2\varrho +m^3 \right) 
\left( 1-\frac{m}{2\alpha \varrho}  \right)^{ \frac{2(\alpha -1)(\alpha 
+2)}{\alpha} +2\alpha +2}  \right\}^{-1} \,.
\end{eqnarray}
\end{widetext}
Since $\alpha >1$ the Ricci scalar diverges as $\varrho \rightarrow \frac{m}{2\alpha 
}$.  Using  eq.~(\ref{isotropicradius}), it is seen that this value 
of the isotropic  radius corresponds to  $\tilde{r}=2m/\alpha$ and 
(using eq.~(\ref{arealradius})) to the  areal radius $r =0$. 
Therefore,  $r \rightarrow 0$  denotes  a central singularity, 
which is a strong one in the sense of Tipler's classification 
\cite{Tipler} because the area of the  2-spheres orbits of 
symmetry vanishes as $r \rightarrow 0$:  an object falling 
onto $r=0$  will be crushed to zero volume.


\end{document}